\documentclass[aps,prl,superscriptaddress,twocolumn]{revtex4-1}
\usepackage{lineno,hyperref}
\usepackage{epstopdf}
\usepackage{epsfig}
\usepackage{epstopdf}
\usepackage{graphicx}
\usepackage{bmpsize}
\usepackage{amsfonts}

\newcommand{\ben}{\begin{eqnarray}}
\newcommand{\een}{\end{eqnarray}}

\newcommand{\bef}{\begin{figure}[h!bt]\centering}
\newcommand{\eef}{\end{figure}}
\newcommand{\bet}{\begin{table}[hbt]\centering}
\newcommand{\eet}{\end{table}}

\begin{document}
\title{Structural and magnetic phase transitions in Ca$_{0.73}$La$_{0.27}$FeAs$_2$ with electron overdoped FeAs layers }
\author{Shan Jiang}
\email{These authors contribute equally.}
\affiliation{Department of Physics and Astronomy and California NanoSystems Institute, University of California, Los Angeles, CA 90095, USA}

\author{Chang Liu$^*$}
\affiliation{Department of Physics, South University of Science and Technology of China, Shenzhen, Guangdong 518055, China}

\author{Huibo Cao}
\affiliation{Quantum Condensed Matter Division, Oak Ridge National Laboratory, Oak Ridge, TK 37831, USA}

\author{Turan Birol}
\affiliation{Department of Physics and Astronomy, Rutgers University, Piscataway, NJ 08854, USA}

\author{Jared M. Allred}
\affiliation{Materials Science Division, Argonne National Laboratory, Argonne, IL 60439-4845, USA}

\author{Wei Tian}
\affiliation{Quantum Condensed Matter Division, Oak Ridge National Laboratory, Oak Ridge, TK 37831, USA}

\author{Lian Liu}
\affiliation{Department of Physics, Columbia University, New York, NY 10027, USA}

\author{Kyuil Cho}
\affiliation{Ames laboratory and Department of Physics and Astronomy, Iowa State University, Ames, IA 50011, USA}

\author{Matthew J. Krogstad}
\affiliation{Materials Science Division, Argonne National Laboratory, Argonne, IL 60439-4845, USA}
\affiliation{Physics Department, Northern Illinois University, DeKalb, IL 60115, USA}

\author{Jie Ma}
\affiliation{Quantum Condensed Matter Division, Oak Ridge National Laboratory, Oak Ridge, TK 37831, USA}

\author{Keith M. Taddei}
\affiliation{Materials Science Division, Argonne National Laboratory, Argonne, IL 60439-4845, USA}
\affiliation{Physics Department, Northern Illinois University, DeKalb, IL 60115, USA}

\author{Makariy A. Tanatar}
\affiliation{Ames laboratory and Department of Physics and Astronomy, Iowa State University, Ames, IA 50011, USA}

\author{Moritz Hoesch}
\affiliation{Diamond Light Source, Harwell Campus, Didcot OX11 0DE, United Kingdom}

\author{Ruslan Prozorov}
\affiliation{Ames laboratory and Department of Physics and Astronomy, Iowa State University, Ames, IA 50011, USA}

\author{Stephan Rosenkranz}
\affiliation{Materials Science Division, Argonne National Laboratory, Argonne, IL 60439-4845, USA}

\author{Yasutomo J. Uemura}
\affiliation{Department of Physics, Columbia University, New York, NY 10027, USA}

\author{Gabriel Kotliar}
\affiliation{Department of Physics and Astronomy, Rutgers University, Piscataway, NJ 08854, USA}

\author{Ni Ni}
\email{Corresponding author: nini@physics.ucla.edu}
\affiliation{Department of Physics and Astronomy and California NanoSystems Institute, University of California, Los Angeles, CA 90095, USA}


\begin{abstract}
We report a study of the Ca$_{0.73}$La$_{0.27}$FeAs$_2$ single crystals. We unravel a monoclinic to triclinic phase transition at 58 K, and a paramagnetic to stripe antiferromagnetic (AFM) phase transition at 54 K, below which spins order 45$^\circ$ away from the stripe direction. Furthermore, we demonstrate this material is substantially structurally untwinned at ambient pressure with the formation of spin rotation walls (S-walls). Finally, in addition to the central-hole and corner-electron Fermi pockets usually appearing in Fe pnictide superconductors, angle-resolved photoemission (ARPES) measurements resolve a Fermiology where an extra electron pocket of mainly As chain character exists at the Brillouin zone edge.
\end{abstract}

\maketitle

  \textit{Introduction.} Both cuprates and Fe-based superconductors, the two known high T$_c$ superconducting families, show rich emergent phenomena near the superconductivity (SC) \cite{1111}. To understand the mechanism of unconventional SC, it is crucial to unravel the nature of these emergent orders. The newly discovered 112 Fe pnictide superconductor (FPS), Ca$_{1-x}$La$_{x}$FeAs$_2$ (CaLa112), shows SC up to 42 K, the highest bulk T$_c$ among all nonoxide FPS \cite{japan112b}. It crystalizes in the monoclinic lattice as a derivative of the HfCuSi$_2$ structure \cite{112a,112b,112c} with the presence of As chains in the spacer layers. Being an exceptional FPS where the global $C_4$ rotational symmetry is broken even at room temperature (Fig. 1(a)), it is important to extract the similarities and differences between CaLa112 and other FPS so that critical ingredients in inducing SC in FPS can be filtered. Efforts have been made to answer whether these nontrivial As chains result in obvious distinctions in the physical properties. Metallic spacer layers are suggested and a fast-dispersing band arising from As chains is observed \cite{gabi,DFT1, ARPESli}. However, until now, no systematic experimental study on this system has been performed to understand the competing emergent orders in these low symmetry systems, which is subject of this paper.

  \textit{The ``parent" phase of CaLa112.} Single crystals of Ca$_{0.73}$La$_{0.27}$FeAs$_2$ were grown using the self-flux method. CaAs, LaAs, FeAs precursors and As powder were ground and mixed thoroughly at the ratio of 1.3:0.5:1:0.7. The mixed powder was then pressed into a pellet, loaded into an Al crucible and sealed into a quartz tube under 1/3 argon atmosphere pressure. The ampule stayed at 1100$^\circ$C for 72 hours, then slowly cooled down to 875$^\circ$C at a rate of 2$^\circ$C/h, followed by water quenching. The partially melted pellet was then removed from the crucible and rinsed by deionized water to get rid of the flux. Silver-shining plate-like single crystals up to 2 mm long were obtained.
Thermodynamic and transport properties were measured using {\it PPMS Dynocool}
and {\it MMPS3} systems from Quantum Design. The elemental analysis was made on several single crystals using wavelength dispersive spectroscopy (WDS) in a JEOL JXA-8200 WD/ED combined microanalyzer. The WDS shows the average La concentration in these single crystals is 0.266(9). For simplicity, we write the chemical formula as Ca$_{0.73}$La$_{0.27}$FeAs$_2$. We have identified Ca$_{0.73}$La$_{0.27}$FeAs$_2$ as the ``parent" compound of CaLa112 and demonstrated it is substantially structurally untwinned at ambient pressure and characterized by the presence of metallic spacer layers.
\begin{figure*}
  \centering
  \includegraphics[width=6.8in]{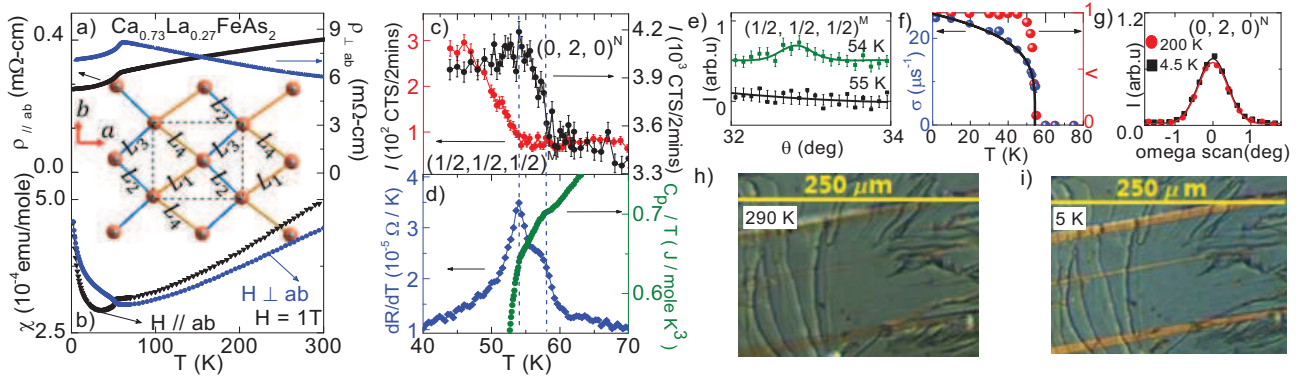}
   \caption{Ca$_{0.73}$La$_{0.27}$FeAs$_2$: (a) Electric resistivity $\rho_{//ab}$ ($I//ab$) and $\rho_{\perp ab}$ ($I\perp ab$) vs. $T$. Inset: The top view of the Fe and spacer As sublattices. The structure subtlety is exaggerated. Orange ball: Fe. Grey ball: As in the spacer layers. The orange and blue lines indicate Fe-Fe bond with bond length of $L_1$,$L_2$, $L_3$ and $L_4$. $L_1$+$L_3$=$L_2$+$L_4$ if $\gamma=90^\circ$. The dash lines enclose the unit cell. (b)Susceptibility $\chi_{//ab}$ and $\chi_{\perp ab}$ vs. $T$. (c) The neutron intensity of the nuclear (0 2 0)$^N$ and the magnetic (1/2 1/2 1/2)$^M$ peaks vs. $T$. (d) Heat capacity $C_p$/T and $d\rho_{//ab}/dT$ vs. $T$. (e) The neutron intensity of the (1/2 1/2 1/2)$^M$ peak at 55 K and 54 K with offset. (f) The magnetically ordered volume fraction $V$ and transverse relaxation rate $\sigma$ in ZF$\mu$SR asymmetry spectra vs. $T$. (g) The neutron intensity of the (0 2 0)$^N$ peak at 200 K and 4.5 K. (h) The polarized optical image at 290 K. i) The polarized optical image at 5 K. }
  \label{fig:Fig1}
  \end{figure*}

The transport and thermodynamic properties of Ca$_{0.73}$La$_{0.27}$FeAs$_2$ are summarized in Fig. 1(a)-(b). The anomalies in resistivity and magnetization data around 60 K are reminiscent of the ones observed in other magnetic FPS, which are associated with the structural/magnetic ordering \cite{dai}. Gradual increase of the inter-plane resistivity, $\rho \bot ab$, is
reminescent of the behaviour found in 122 family of compounds \cite{rc1, tanatarc}, where it was assigned to pseudo-gap formation. The ratio
of $\rho ^{\perp ab}/\rho ^{// ab}$ increases from 15 to 30 upon cooling. If we assume no low-energy spin excitation exists akin to other magnetic FPS \cite{gap}, via fitting the low temperature $C_p$ by $C_p=\gamma T+\beta T^3$, we find $\gamma$ is 12.2 mJ/mole K$^2$ and the Debye temperature $\theta_{D}$ = 346 K, which is much higher than $\theta_{D}$ = 260 K in BaFe$_2$As$_2$ \cite{sefat,hardy}, pointing to a stiffer lattice in CaLa112. To reveal the nature of the anomalies observed in Fig. 1(a)-(b), neutron diffraction data were taken on Ca$_{0.73}$La$_{0.27}$FeAs$_2$. The (0 2 0)$^N$ nuclear Bragg peak intensity, measured on the single crystal I with multiple growth domains \cite{suppl}, increases below 58 K (Fig. 1(c)), signaling a structural phase transition. Magnetic Bragg peaks appear at lower temperature. As a representative, the (1/2 1/2 1/2)$^M$ magnetic Bragg peak is shown in Fig. 1(e). It is absent at 55 K but clearly present at 54 K. These observations evidence a structural phase transition at $T_s$=58 K and an antiferromagnetic (AFM) phase transition at $T_m$=54 K, consistent with the two-kink feature in the $C_p/T$ and d$\rho/$d$T$ (Fig. 1(d)). The temperature evolution of the magnetically ordered volume fraction $V$ was determined from $\mu$SR data, as shown in Fig. 1(f) \cite{suppl}. For reference we also show temperature evolution of the transverse relaxation rate, $\sigma$, which is proportional to the local
magnetic moment. The fact that $\sigma$ increases much slower below $T_m$ than $V$ provides good evidence for homogeneous magnetic order in the sample. Due to the existence of $T_s$/$T_m$ in Ca$_{0.73}$La$_{0.27}$FeAs$_2$, we call it as the ``parent" compound of CaLa112. The 42 K SC in Ca$_{0.82}$La$_{0.18}$FeAs$_2$ arises from hole doping through Ca substitution on the La sites.

\begin{figure*}
  \centering
  \includegraphics[width=6.4in]{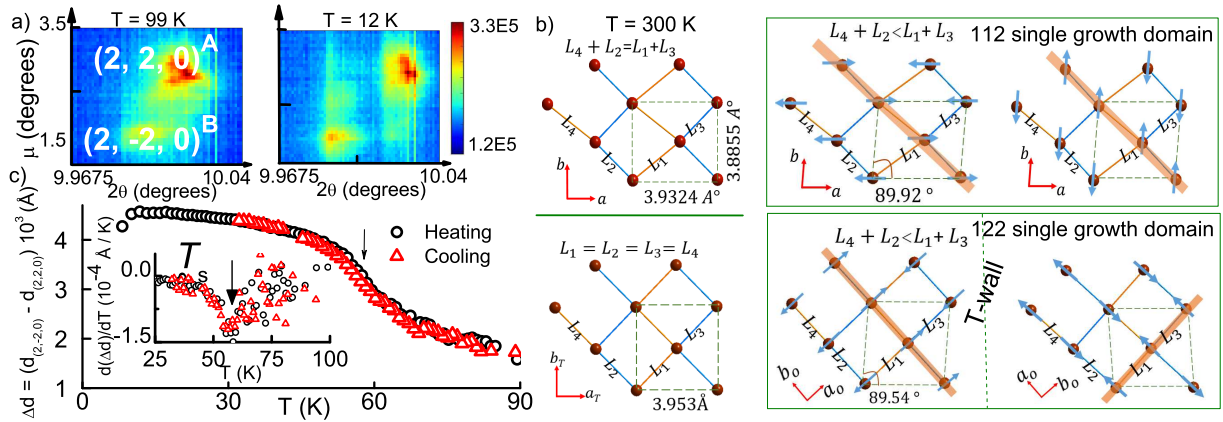}
  \caption{(a) The synchrotron x-ray $\mu$ vs. 2$\theta$ diffractograms of the (2 2 0)$^A$ and (2 -2 0)$^B$ peaks from the growth domain A and B of Ca$_{0.73}$La$_{0.27}$FeAs$_2$, respectively. $\mu$ is a rotation of the crystal along an axis perpendicular to the x-ray beam. (b) The cartoon comparison of the magnetic and crystal structures between Ca$_{0.73}$La$_{0.27}$FeAs$_2$ and BaFe$_2$As$_2$ in a single growth domain. Orange ball: Fe. Blue arrow: the spin direction. Orange ribbon: the spin stripe along which the spins order in parallel. The dash lines enclose the 2-Fe cell. (c) Splitting of the two reflections as determined by Gaussian fits. Inset: $d\Delta d/dT$ vs. $T$ .}
  \label{fig:Figa}
\end{figure*}

\textit{Lack of substantial twinning below $T_s$.} A single crystal II of Ca$_{0.73}$La$_{0.27}$FeAs$_2$ with only one growth domain was measured at HB-3A (Fig. 1(g)) \cite{suppl, hb3a}. Within the measurements' resolution, its (0 2 0)$^N$ nuclear peak shows no broadening caused by extinction effect at 200 K and 4.5 K. This is dramatically different from the BaFe$_2$As$_2$ at ambient pressure \cite{wilson}. Inside BaFe$_2$As$_2$, the (2 0 0)$^N$ peak broadens below $T_s$ because of the formation of structural twinning walls (T-walls), across which, the spin pattern, spin orientation and crystalline axis rotate 90$^\circ$ \cite{NiO, tanatar3}. Instead, Fig. 1(g) is reminiscent of the detwinned BaFe$_2$As$_2$ under 0.7 MPa \cite{wilson}, where T-walls are wiped off and no (2 0 0)$^N$ peak broadening appears. This suggests that either the sample is substantially untwinned or the twinning below $T_s$ is too weak to cause detectable broadening. Figure 1(h)-(i) show the polarized optical images taken on Ca$_{0.73}$La$_{0.27}$FeAs$_2$. At 290 K, the growth domains are bordered by the bright lines. Unlike the extra $\mu m$-sized structural domains observed in BaFe$_2$As$_2$ below $T_s$ \cite{tanatar3}, none is observed in Ca$_{0.73}$La$_{0.27}$FeAs$_2$ at 5 K. However, since good surface condition is critical to image the T-walls using this technique, we can not preclude the T-walls formation. To examine if Ca$_{0.73}$La$_{0.27}$FeAs$_2$ is indeed substantially untwinned, a single crystal III with two growth domains was picked for the synchrotron x-ray diffraction measurement. The $a$ axis in the growth domain A and B are rotated by 90$^\circ$ relative to each other \cite{suppl}. Figure 2(a) shows the synchrotron x-ray intensity of the (2 2 0)$^A$ and (2 -2 0)$^B$ peaks of the growth domain A and B, respectively \cite{suppl}. The slight separation of these two peaks even at 99 K comes from tiny differences in orientation/position between growth domains A and B. The profiles of both peaks include a 2$^\circ$ shoulder along the $\mu$ axis. The shoulders are visible at 12 K but overlap at 99 K since a projection along a 3rd, orthogonal axis is used to create the plot. In the non-reduced data the (2 2 0)$^A$ and (2 -2 0)$^B$ can be resolved at all temperatures, which unambiguously shows that neither peak splits below 58 K \cite{suppl}. This is in sharp contrast with the 122, where the (2, 2, 0) synchrotron x-ray peak splits into two/four blobs with similar brightness below $T_s$ because of the structural twinning \cite{tanatar3}. A conservative estimation points to 95\% of each growth domain being untwinned in single crystal III. This feature may be related to the As chains in the crystal, which make the T-walls formation energetically unfavorable in Ca$_{0.73}$La$_{0.27}$FeAs$_2$, being consistent with the stiffer lattice suggested by the specific heat measurement.

\textit{The magnetic structure.} We determined the magnetic structure of Ca$_{0.73}$La$_{0.27}$FeAs$_2$ based on 13 effective magnetic reflections of the single crystal II \cite{suppl}. Figure 2(b) shows the comparison between it and BaFe$_2$As$_2$ in a single growth domain. Now we focus mainly on the magnetic structure and defer the discussion of the crystal structure later. Since synchrotron x-ray diffraction shows that Ca$_{0.73}$La$_{0.27}$FeAs$_2$ is substantially untwinned, no T-walls are considered in the neutron data fitting. We found that a reasonable fit requires two equal volumes with different spin orientations (blue arrows) either along $a$ or $b$ axis \cite{suppl}. Combining the fact that the sample is substantially untwinned, this suggests the formation of the spin rotation walls (S-walls), across which the spin pattern and crystallographic axis stay the same but the spin orientations rotate \cite{NiO}. Although the detailed magnetic structure of this compound is unique with the antiparallel spins being off-head to each other instead of head-to-head, it has the same AFM stripe pattern with the wave vector of (1, 0) in the 1-Fe cell akin to other FPS \cite{dai, huang}, being consistent with our DMFT calculations. The existence of ``magnetic domains" with the easy axis 45$^\circ$ or 135$^\circ$ away from the stripe direction suggests that the magnetic anisotropy energy of these two is so close to each other that they can practically coexist, consistent with the DFT anisotropy energy of $\sim$0.1 meV / Fe. Our DMFT calculations predict an ordered moment of 1.0 $\mu_B$ / Fe, which agrees well with the experimental value, 1.08(3) $\mu_B$ / Fe.

\textit{Monoclinic to triclinic structural phase transition.} We now discuss the nature of the structural phase transition illustrated in Fig. 2(b). In all known magnetic FPS, because of the magneto-elastic coupling, the onset of the (1, 0) stripe magnetic order breaks the 4-fold rotational symmetry and leads to a tetragonal to orthorhombic phase transition. As a result, the 2-Fe cell enclosed by dash lines in the lower panel of Fig. 2(b) distorts from an exact square to a rhombus with the short diagonal along the stripe direction \cite{jared}. Since the magnetic wave vector of Ca$_{0.73}$La$_{0.27}$FeAs$_2$ is the same as those in other magnetic FPS, we expect similar type of magneto-elastic coupling, which breaks the only symmetry element of $P2_1$ and reduces it to triclinic $P1$. This leads to $\gamma\neq 90^\circ $. Consequently, the 2-Fe cell distorts from a rectangle into a parallelogram (upper panel in Fig. 2(b)) and ($L_1$+$L_3$) is no longer equal to ($L_2$+$L_4$) below $T_s$. Figure. 2(c) shows that upon cooling, the difference in $d$ spacing between these two reflections in Fig. 2(a) monotonically increases. A sharp kink at 58 K appears in d$\Delta d/$d$T$. Assuming $\alpha \sim 90^\circ$, the $\Delta d$ gives a cell with $\gamma$ = 89.92$^\circ$ and ($L_1$+$L_3$)-($L_2$+$L_4$) = 0.007(4)$\AA$ at 10 K \cite{suppl}, suggesting weak spin-orbit coupling.

\begin{figure*}
  \centering
  \includegraphics[width=6.8in]{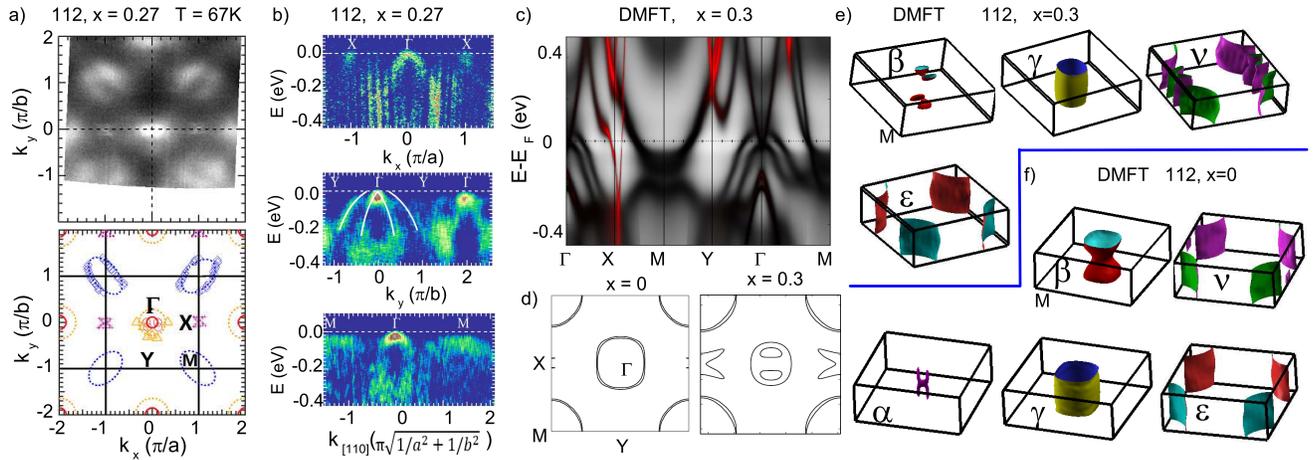}
  \caption{(a)The 2D contour of the ARPES Fermi surface (FS) of Ca$_{0.73}$La$_{0.27}$FeAs$_2$ at $K_z \sim \pi/c$ in the 2 Fe/cell representation. The red and orange circles: the two hole pockets at the center $\Gamma$ point. The blue ovals: the electron pockets at the corner M point. The purple lines: the extra electron pocket arising from the As chains at the X point. (b) The 2nd derivative of ARPES $k-E$ maps. Two hole pockets at $\Gamma$ points can be clearly identified in the Y-$\Gamma$ cut. (c) The spectral function A(k, $\omega$) of Ca$_{0.7}$La$_{0.3}$FeAs$_2$ from DMFT. The red color represents the projection of the orbital character onto the in-plane $p$ orbitals of the As-chain atoms. (d) The 2D contour of the DMFT FS of CaLa112 ($x=0$ and 0.3) at $K_z \sim \pi/c$.
  (e) and (f): The 3D DMFT FS of the CaLa112($x=0.3$) (e) and the CaLa112($x=0.0$) (f) in the 2Fe/cell representation. }
  \label{fig:DMFTfermi}
\end{figure*}

\textit{Why is the electron overdoped FeAs layer magnetic in CaLa112 ?} In Ca$_{0.73}$La$_{0.27}$FeAs$_2$, the nearest neighbor As-As distance is 2.56(1) ${\AA}$. It is much shorter than the critical distance 3.00 ${\AA}$ where As-As start bonding \cite{pj}, but longer than the As-As single bond distance \cite{asbook}, 2.46${\AA}$, suggesting the bond order is slightly smaller than 1. Therefore we can model the effective charges here as Ca$^{2+}_{0.73}$La$^{3+}_{0.27}$[FeAs]$^{(1.27-\delta)-}$As$^{(1+\delta)-}$ ($\delta>0$). This indicates the FeAs layer is doped by 0.27-$\delta$ electrons/Fe. ARPES data and DMFT calculations in the paramagnetic phase provide us quantitative understanding of the electronic structure. ARPES of Ca$_{0.73}$La$_{0.27}$FeAs$_2$ shows a unique Fermiology among all FPS (Fig. 3(a)-(b)). Using the $s$ geometry (electric fields out of the emission plane), at 67 K, ARPES resolves two hole pockets at the Brillouin zone center $\Gamma$ and one oval-like electron pocket at the corner M akin to the other magnetic FPS \cite{suppl}. Interestingly, an extra electron pocket, which has never been observed before, appears at the Brillouin zone edge (X point). This is qualitatively consistent with the DMFT calculation (Fig. 3(c)-(e)). In addition to the two hole pockets ($\beta$ and $\gamma$) at $\Gamma$ and two similar-sized electron pockets at M with only FeAs layer character, DMFT also reveals one extra electron pocket at X with only As chain character (Fig. 3(c)). By calculating the volume difference between the Fermi pockets at $\gamma$ and M, the DMFT calculation indicates the FeAs layer is doped by 0.17 e/Fe. Since the ARPES $k_z$ dispersion has not been measured, assuming all pockets are 2D-like, a rough estimation of the ARPES Fermi volume suggests a doping level of $\sim$ 0.2 e/Fe. Comparing with the prototype electron doped Ba122 \cite{chang}, this value places Ca$_{0.73}$La$_{0.27}$FeAs$_2$ as electron overdoped. This seems inconsistent with the current consensus that the parent compound of a FPS is a semimetal with equal numbers of holes and electrons. A closer look into the Fermi surface (FS) shows that, despite of the overdoped nature, a reasonable FS nesting much stronger than the one in electron overdoped Ba122 \cite{chang}, survives in this ``parent" compound. This highlights the important role of FS nesting in inducing structural/magnetic instabilities. On the other hand, Fig. 3(d) shows that the DMFT FS nesting is enhanced upon decreasing $x$ (hole doping). Since experimentally the $T_s$/$T_m$ are suppressed rather than enhanced with decreasing $x$, this suggests that the superexchange interaction also plays a role in causing these instabilities. Therefore, this is one strong piece of evidence of the dual itinerant and localized nature of magnetism in FPS \cite{daireview}. Both the comparison of the ARPES FS between the ``parent" and SC CaLa112 (nominal $x$=0.1, real $x$=0.18, $T_c$ = 42 K) \cite{ARPESli,zhou} and the comparison of the DMFT FS between $x=0.3$ and $x=0$ CaLa112 (Fig. 3(e)(f)) reveal the As chains deeply affect the doping mechanism. With Ca doping, part of the holes create an extra 3D hole pocket ($\alpha$ pocket) at $\Gamma$ with a mixed Fe and As-chain nature \cite{ARPESli}, part of them fill the X electron pocket, and part of them distribute to the rest of the pockets.

A tunable FPS with metallic layers can shed lights on the role of interlayer coupling on the interplay of magnetism and SC in FPS. Recently, 10-4-8 FPS family and Ba$_2$Ti$_2$Fe$_2$As$_4$O have been found to be self doped with metallic spacer layers \cite{hole,fengdl}. However, there is no good control on the extent of self-doping. Therefore, the CaLa112 system is more promising for the systematic study of the impact of metallic layer in FPS. What's more, since the $C_4$ rotational symmetry is already broken even at room temperature, CaLa112 raises new opportunity to study the electronic nematicity, which lowers the rotational symmetry but keeps the translational symmetry and manifests as the in-plane electronic anisotropy of the 1-Fe cell \cite{fisher, tanatar2, impurity, impurity2, zxshen, pengcheng}.

\textit{Summary.} In conclusion, Ca$_{0.73}$La$_{0.27}$FeAs$_2$ with electron overdoped FeAs layer, is the ``parent" compound of the CaLa112 FPS. The magnetic CaLa112 is substantially untwinned with S-walls under ambient pressure. Furthermore, while the central-hole and corner-electron Fermi pockets appear with reasonable nesting, both the ARPES and DMFT unravel an extra electron pocket at the Brillouin zone edge originating from the As chains, establishing the As chains actively participating in the doping mechanism. These characteristics make this material a great platform to study the roles of electronic nematicity and metallic spacer layers in FPS.

{\it Acknowledgments.}Work at UCLA was supported by the U.S. Department of Energy (DOE), Office of Science, Office of Basic Energy Sciences (BES) under Award Number DE-SC0011978. Work at ORNL$^\prime$s High Flux Isotope Reactor and at ANL's Advanced Photon Source was sponsored by the Scientific User Facilities Division, BES, DOE. Work at Argonne and Ames was supported by the Materials Sciences and Engineering Division, BES, DOE. Ames Lab is operated for the DOE by ISU under Contract No. DE-AC02-07CH11358. Work at Columbia and TRIUMF was supported by the NSF DMREF DMR-1436095, PIRE project IIA 0968226 and DMR-1105961. Work at Rutgers was supported by the NSF DMREF DMR-1435918.

\end{document}